\documentclass[twocolumn]{article}
\usepackage{tikz}
\usepackage{epsfig}
\usepackage{booktabs}
\usepackage{inputenc}
\usepackage[english]{babel}
\usepackage{amsmath}
\usepackage{mathtools, cuted}
\usepackage{amsfonts}
\usepackage{amssymb}
\usepackage{graphicx}
\usepackage{listings}
\usepackage{rotating}
\usepackage{authblk}
\usepackage{placeins}
\usepackage[breaklinks=true]{hyperref}

\usepackage{multirow}

\begin{document}

\title{Effects of Early Warning Emails on Student Performance}
%
%
%
%
%

\author[a]{Jens Klenke}
\author[a]{Till Massing\thanks{Corresponding author. Email:
till.massing@uni-due.de}}
\author[a]{Natalie Reckmann}
\author[a]{Janine Langerbein}
\author[b]{Benjamin Otto\thanks{For more information regarding JACK contact this author. Email: benjamin.otto@uni-due.de}}
\author[b]{Michael Goedicke}
\author[a]{Christoph Hanck}
\affil[a]{\footnotesize Faculty of Business Administration and Economics, University of
Duisburg-Essen, Universit{\"{a}}tsstr.~12, 45117 Essen, Germany}
\affil[b]{\footnotesize paluno - The Ruhr Institute for Software Technology,
University of Duisburg-Essen, Gerlingstr.~16, 45127 Essen, Germany}



\maketitle
\begin{abstract}
	Individual support for students in large university courses is often difficult. To give students feedback regarding their current level of learning at an early stage, we have implemented a warning system that is intended to motivate students to study more intensively before the final exam. For that purpose, we use learning data from an e-assessment platform for an introductory mathematical statistics course to predict the probability of passing the final exam for each student. Subsequently, we sent a warning email to students with a low predicted probability of passing the exam. Using a regression discontinuity design (RDD), we detect a positive but imprecisely estimated effect of this treatment. Our results suggest that a single such email is only a weak signal – in particular when they already receive other feedback via midterms, in-class quizzes, etc. – to nudge students to study more intensively.
\end{abstract}

%


\section{\uppercase{Introduction}}
\label{sec:intro}

Introductory courses at universities are often attended by a large number of students. Many such courses are in the first academic year, with students struggling with increased anonymity and distributing their workload. It is not easy for lecturers to support each student individually. One can, however, inform students if their performance throughout class indicates a high probability of them failing the final exam. \cite{Massing2018a} show that it is possible to predict students’ outcome in the final exam at a relatively early stage of an introductory statistics course. These early-warning procedures may help students to better assess their level of proficiency.

For such prediction, it is attractive to use data from e-learning platforms to gain insights into the students' learning behavior. In this paper, we use a logit model to predict the probability of a student passing the final exam. For model building, we use data from a previous cohort of the course. To establish the early-warning system, we send emails to students of the current cohort of the course with a low predicted probability of passing the final exam. This intervention is to serve as a wake-up call for students who may improperly assess their level of proficiency. 
We investigate the effectiveness of said warning emails by using a regression discontinuity design (RDD). We find that the emails have a positive but imprecisely estimated effect on the students' performance in the final exam.

The remainder of this paper is organized as follows: Section \ref{sec:method} describes the statistics course investigated in the study. Section \ref{sec:related} provides a brief overview of related work. We present the available data and the models used in section \ref{subsec:data}. Section \ref{sec:analysis} discusses the empirical results. Section \ref{sec:conclusion} concludes.

\section{\uppercase{Course Description}}\label{sec:method}

\begin{figure*}[]
	\resizebox{\textwidth}{!}{%
			\input{timeline_plot.tex}
			}
  		\caption{Timeline for the key events in the 2019 summer term course \textit{Inferential Statistics} (treatment cohort). The shaded area indicates the period after treatment. There were $57$ days between the warning email and the first opportunity to take the exam and $113$ days between the warning email and the second opportunity.
  		} 
	\label{fig:timeline}
\end{figure*}

This section outlines the structure of the \textit{Inferential Statistics} course at the University of Duisburg-Essen in 2019, in which students at risk received a warning email. The course is compulsory for several business and economics programs and teachers' training, and hence we gathered information on 802 students from JACK and Moodle. Of these 802 students, 337 took an exam at the end of the semester.\footnote{Note that the 802 students' data in our platforms do not imply that all of them actively followed the course. JACK and Moodle are open to many students, not only those who want or need to participate in the class.}  

The course consisted of a weekly 2-hour lecture, which introduced concepts, and a 2-hour exercise, which presented explanatory exercises and problems. We conducted at least one Kahoot!~game in every lecture and exercise. These games were short quizzes related to the subject of the current class. Students were able to earn up to $2$ bonus points. Both classes were held classically in front of the auditorium. Because these classes have many students, they are limited in their ability to address students' varying learning speeds and individual questions. To overcome this and to encourage self-reliant learning and supporting students who had difficulties attending classes, all homework was offered on the e-assessment system JACK, where the correctness of students' answers is automatically assessed.

In addition to classical fill-in and multiple-choice exercises, the course also introduces the statistics software R. Individual learning success is supported by offering specific automated feedback and optional hints. Students were able to ask additional questions in a Moodle help forum.

We offered five online tests using JACK to encourage students to study continuously during the semester and not only in the weeks prior to the exams. These tests lasted 40 minutes at fixed time slots. These summative assessments allowed students to assess their individual state of knowledge during the lecture. Students did not need to participate in online tests to take the final exam at the end of the course.
Instead, we offered up to 10 bonus points to encourage participation. The bonus points were only added to final exam points if students passed the exam without the bonus. Students may earn up to 60 points in the exam. We provide more detail on the data in section \ref{subsec:data}.

Kahoot!~games, exercises on JACK, and online tests already provide students with some feedback during the semester. 

Before the last online test of the previous cohort in 2017\footnote{The course is jointly offered by two chairs, and therefore held on a rotating basis. Hence, the course is only comparable every two years.}, the points students reached in JACK exercises and the previous online tests were analyzed to predict individual probabilities of passing the final exams. In the current cohort in 2019, we split students into three groups according to the trained model: students with a high probability of passing the exam, one group with a moderate probability and the last group with a low probability of passing the exam. The students in the last two groups received a warning email, which was formulated more strictly for those with a low probability of passing the course. 

Students received the warning email shortly after the third online test on June $6^{th}$. The first possibility to take the final exam was on August $2^{nd}$. The second and last possibility to take the final exam was on September $27^{th}$. Students who did not pass the final exam on the first attempt were allowed to retake the final exam. Section \ref{sec:analysis} will analyze the effectiveness of the email regarding the passing probability of the final exam.

The final exams were also held electronically. While online tests during the semester could be solved at home with open books, the final exams were offered exclusively at university PC pools and proctored by academic staff. Students can only retake an exam if they failed or did not take the previous one (so that students can pass at most once) but can fail several times.\footnote{Students obtain 6 “malus points” for each failed exam, of which they may collect up to 180 during their whole bachelor's program. This has the side effect that showing up unprepared and hence failing (many) single exams has limited consequences. Predicatively, this relates into relatively low passing rates in our program.} The maximum number of points a student achieves in an exam (over both exams per semester) determines the final grade. We denote the corresponding exam as the final exam.

We summarize the timeline of the main events of the course in Figure~\ref{fig:timeline}. 

\section{\uppercase{Related work}}\label{sec:related}

Students' general interest and participation are among the strongest factors for successful studies \cite{koccak2021,schiefele1992}. \cite{Sosa2011} show in a meta-study that the additional use of e-assessment in traditional face-to-face courses positively affects student success. \cite{Massing2018a} measure learning effort and learning success via the total number of (correct) entries on the e-assessment platform JACK during the course. They can show that this positively influences the final exam grade. 

The literature has identified a number of important predictors. \cite{Gray2014} show the importance of socioeconomic and psychometric variables as well as pre-high school grades, which may vary across countries \cite{Oskouei2014}. 
Variables that arise after admission to the program, such as credits earned, degree of exam participation, and exam success rate in previous courses, are also related to student success \cite{Baars2017}. \cite{Macfadyen2010,Wolff2013} analyze student activity in learning management systems to accurately predict student performance. \cite{Burgos2018,Huang2013,Massing2018b} use the activity in e-learning frameworks and the results of midterm exams to predict student success in the final exam, which is particularly useful as these predictors are related to the exam under consideration. 

Some researchers have already worked on identifying students at risk in higher education. \cite{akccapinar2019b,akccapinar2019,chen2020,chung2019,lu2018}
study the identification of students at risk in different contexts of e-learning systems. They all show that it is possible to achieve high accuracy in predicting student success early in the semester. Furthermore, \cite{baneres2020} implemented an early warning system. However, they did not analyze the effect of the system on students’ performance. 

To summarize, data on student learning activity can be used for the implementation of early warning systems. Purdue University (West Lafayette), Indiana, developed the early warning system Course Signals, see \cite{Arnold2010}. Email notifications and signal lights (red, yellow, and green) inform students of their learning status. \cite{Arnold2012} analyze the retention and performance outcomes achieved since the introduction of Course Signals. They find that using this early warning system substantially affects students' grades and retention behavior. \cite{csahin2019} developed, based on learning analytics, an intervention engine called the Intelligent Intervention System (In2S). In this system, students see signal lights for each assessment task, representing a direct intervention in the classroom. The system uses gamification elements such as a leaderboard, badges, and notifications as an additional motivational intervention. Learners who use In2S emphasize the system's usefulness.

\cite{Iver2019} examined the effects of a ninth-grade early warning system on student attendance and percentage of credits earned in ninth grade. They could not find a statistically significant impact of the intervention. \cite{Edmunds2002,stanfield2008} investigate the effects of different incentives on third or fourth-graders' reading motivation. \cite{Edmunds2002,Iver2019,stanfield2008} find no significant differences in reading motivation between students who received incentives and those who did not.  \cite{Parkin2012} used a range of technical interventions that can encourage the effort level. They can show that online posting of grades, feedback and adaptive grade release significantly improve students' engagement with their feedback.

Most researchers analyze the corresponding early warning system qualitatively via questionnaires using linear regression (OLS). \cite{lauria2013} show that their warning system leads to a higher dropout rate of students receiving a warning. An increased dropout rate  is often viewed negatively. However, they argue that there are also many positive aspects from the perspective of students and instructors, such as a higher focus on other courses, less sunk cost, and fewer students to supervise, which may increase support for other students.

OLS cannot be used in our settings. The issue is that the warning is inherently not assigned randomly but instead based on student performance. Therefore a standard OLS regression of student success on, e.g., variables such as whether a warning was issued would not identify the causal effect of the warning. Such a regression would be confounded with unobserved influences such as general ability or motivation, which both affect the outcome, student success, as well as whether a student receives a warning as the warning is issued to those students, which will be less qualified/motivated on average compared to students who have not received a warning.    

However, regression discontinuity designs (RDD) may be suitable in such settings as the method can isolate the potential effect from other influences.  In this design, there are two groups of individuals, in which one group receives a specific treatment, such as an early warning. A running variable, $W$, gives the individual probability for each student to pass the exam. The value of this running variable lying on either side of a fixed threshold determines the assignment to the two groups. Comparing individuals with values of the running variable below the threshold to those just above can be used to estimate the causal effect of the treatment on a specific outcome. \cite{McEwan2008} use regression discontinuity approaches to estimate the effect of delayed school enrollment on student outcomes, as these students will be similar w.r.t. observed as well as unobserved confounders such as those mentioned below. 

\cite{Angrist1999} use the RDD approach to estimate the effect of class size on test scores. \cite{Jacob2004} studied the effect of remedial education on student achievement. More details on RDD will be provided in the next section. 

\section{\uppercase{Data and Model}}\label{subsec:data}

\begin{table*}[t]
	\center
	\caption{
	Overview of empirical quartiles, mean and standard deviation for the response variable and considered covariates. Exam points describes the points reached in the final exam. Online test is the sum of the first four online tests, while the JACK score describes the score until the warning mail was sent.}
	\begin{tabular}{@{}lccrrrrrrr@{}}
\toprule
 variable    	&	warning	& count	& min		& Q0.25		& median	& mean		& Q0.75		& max		& sd	\\ \midrule
 Exam points	& 0			& 151	& 3.40		& 17.50		& 25.30		& 23.80		& 30.00		& 47.00		& 8.20 \\
 				& 1			& 183	& 0.00		& 8.55		& 15.60		& 16.10		& 23.10		& 39.20		& 9.52\\
 Online test 	& 0			& 191	& 341.00	& 579.00	& 755.00	& 761.00	& 900.00	& 1425.00	& 221.00 \\
 				& 1 		& 607	& 0.00		&	0.00	& 0.00		& 98.50		& 167.00	& 700.00	& 133.00\\	
JACK score 		& 0 		& 191	& 0.00	 	& 2216.00	& 3520.00	& 3799.00	& 5174.00	& 11580.00	& 2206.00 \\
				& 1 		& 425	& 0.00		& 200.00	& 983.00	& 1347.00	& 2033.00	& 8385.00	& 1363.00\\
\bottomrule
	\end{tabular}
	\label{tab:group_overview}
\end{table*}

This section presents the data and model used for the analysis. The raw data is collected from three different sources. First, we collected each student's homework submissions on JACK, where we monitored the exercise ID, student ID, the number of points (on a scale from 0 to 100) and the time stamp with a minute-long precision. The second data source comprises the online tests, whereby the student may earn extra points for their final grade, see section \ref{sec:method}. Until the treatment was assigned, three out of five online tests were conducted. Lastly, the response variable is given by their final exam result. For students' final grades, which consist of the final exam result and earned bonus points, the following grading scheme was applied: very good (``1''), good (``2''), satisfactory (``3''), sufficient to pass (``4''), and failed (``5''). We assigned ``6'' to 465 students who participated in the course but did not take any final exams.\footnote{Postponing exams to later semesters is possible and common in our program.}  This reflects our view that students who did not take any exam were even less prepared than students who failed the exams. 

Over the whole course, JACK registered $175,480$ submissions of homework exercises from students. For each student, we compute the score (JACK score in Table~\ref{tab:group_overview}) as the sum of the points of the latest submission over all subtasks. 
 
To determine who should receive a treatment (warning email), we mainly considered the results from the first three online tests, which had been conducted until then. We used a logit model to predict the probability that a student with these online test results would pass the exam. The model was trained with the data obtained from the same course given two years earlier, see \cite{Massing2018a,Massing2018b}. The predicted probability will serve as our running variable $W$ in the \textit{RDD}, see equation \eqref{eq:rdd_mod} below. These predictions were transformed to an ordinal variable. If the predicted probability of passing the exam was larger than $0.4$, the student was be supplied with  \textit{no} message (0 in Table~\ref{tab:group_overview}), and with less than 0.4 with a warning message (1 in Table~\ref{tab:group_overview}).

However, as the online tests were not mandatory, we further considered the students' general activity during the course and thus modified the decision on whether an email was sent for a subset of the students. By taking into account activity during the course, we eliminate two disadvantages that could arise if we solely built the treatment on the online tests.

First, the online tests were not mandatory, and not all students took the online test -- despite the high incentive to earn extra points for their final grade.\footnote{There are several possible explanations for that. E.g., perhaps some students could not participate due to other commitments since the online tests were held at a fixed date and time.} 

Second, the students were allowed to cooperate during the online tests, although all students were graded individually. This could potentially lead to a student performing well in the online tests, although he did not comprehend the course content that well.

Given our data and treatment design, which were not distributed randomly but rather based on the probability to pass the exam, we use the \textit{RDD} to analyze the effectiveness of our intervention.\footnote{We also investigated alternative modeling approaches like propensity score matching. However, the results were similar, and RDD seems most suitable given the problem.}

The method allows us to compare students around the cutoff point and hence to causally identify a possible treatment effect. Our identifying assumption is that the participants around the cutoff are similar with respect to other (important) properties. Such determinants, for example, include the general or quantitative ability. To distinguish between the different RD designs, first consider
\begin{align}
	\label{eq:rdd_mod}
	Y_i = \beta_0 + \alpha T_i + \beta W_i + u_i
\end{align}
and let
\begin{align}
	T_i  = \begin{cases}
		1 , & W_i \leq c,\\
		0, & W_i > c,
	\end{cases}
\end{align}
where $T_i$ indicates if a student received an email, which is determined by the threshold $c$, in our case 0.4 of the predicted probability to pass the exam $W_i$. $Y_i$ is the sum of points of student $i$ in their (latest) final exam, and $u_i$ is an error term. For the analysis, only students who attended at least one final exam were included ($n = 337$). This design deterministically assigns the treatment, which means that only if $W_i \leq c$ will the student receive the treatment. The  treatment effect is represented by $\alpha$.

The approach sketched above is a \textit{sharp RDD} since the two groups (treatment, no treatment) are perfectly separated by the cutoff. 

However, as explained above, this is not the case in our design as we also wanted to consider the student's activity in the course in our decision. Thus, the groups are no longer \emph{perfectly} separated.

We, therefore, employed an extension of this design, called \textit{fuzzy RDD}. In this case, only the probability of receiving the treatment needs to increase considerably at the cutoff and not from 0 to 1, as in the sharp design. This non-parametric approach estimates a \textit{local average treatment effect (LATE)} $\alpha$ in equation \eqref{eq:2_stage} through an instrumental variable (IV) setting \cite{angristidentification1996}.

More specifically, consider the following model

	\begin{align}
		 \label{eq:2_stage}
								Y_i & = \beta_0+\alpha \ \widehat{T_i}  + \delta_1 W_i + X_i^{\mathrm{T}} \ \pmb{\beta} + u_i	\\
	  							T_i & = \gamma_0 + \gamma_1 \ Z_i + \gamma_2 \ W_i + \nu_i,  \label{eq:1_stage}
	\end{align}

\noindent where equation \eqref{eq:1_stage} represents the first stage of the IV estimation with $T_i$ denoting if a student received the treatment, the instrument $Z_i = 1\left[ W_i \leq c \right]$ indicating if a student is below or above the cutoff of $c = 0.4$ (as in the sharp RDD), $W_i$ remains the predicted probability to pass the exam from the logit model, while $\nu_i$ represents an error term. The fitted values $\widehat{T_i}$ of $T_i$ are inserted into equation \eqref{eq:2_stage}, where $Y_i$ again represents the sum of points of student $i$ in their (latest) final exam. $u_i$ represents the error term, $X_i$ a covariate -- here the sum of points of online tests -- and $\alpha$ the treatment effect.

The following assumptions must be met to identify a potential treatment effect; (i) the running variable $W$ needs to be continuous around the cutoff, see \cite{mccrarymanipulation2008}. If this assumption is not met, it could indicate that participants can manipulate the treatment. Furthermore, the general assumptions for IV estimation must hold. Therefore, (ii) instrument $Z$ only appears in equation \eqref{eq:1_stage} for $T$  and not in equation \eqref{eq:2_stage} for $Y$ (the exclusion restriction). (iii) Instrument $Z$  must be correlated with the endogenous explanatory variable $T$ \cite[pp. 883-885]{Cameron2005}.

We will return to assumption (i) in section \ref{sec:analysis}. The exclusion restriction for the instrument variable $Z$ holds since the variable is only an indicator variable showing whether a student is to the left or right of the cutoff ($c = 0.4$), and the probability of passing the exam is also already included in the second stage of the design. Assumption (iii) is satisfied in an RDD model by the construction of the approach as the instrument is a nonlinear (step) transformation of the running variable \cite{Lee2010}.  

\begin{figure}[]
	\centering
  		\includegraphics[width=0.45\textwidth]{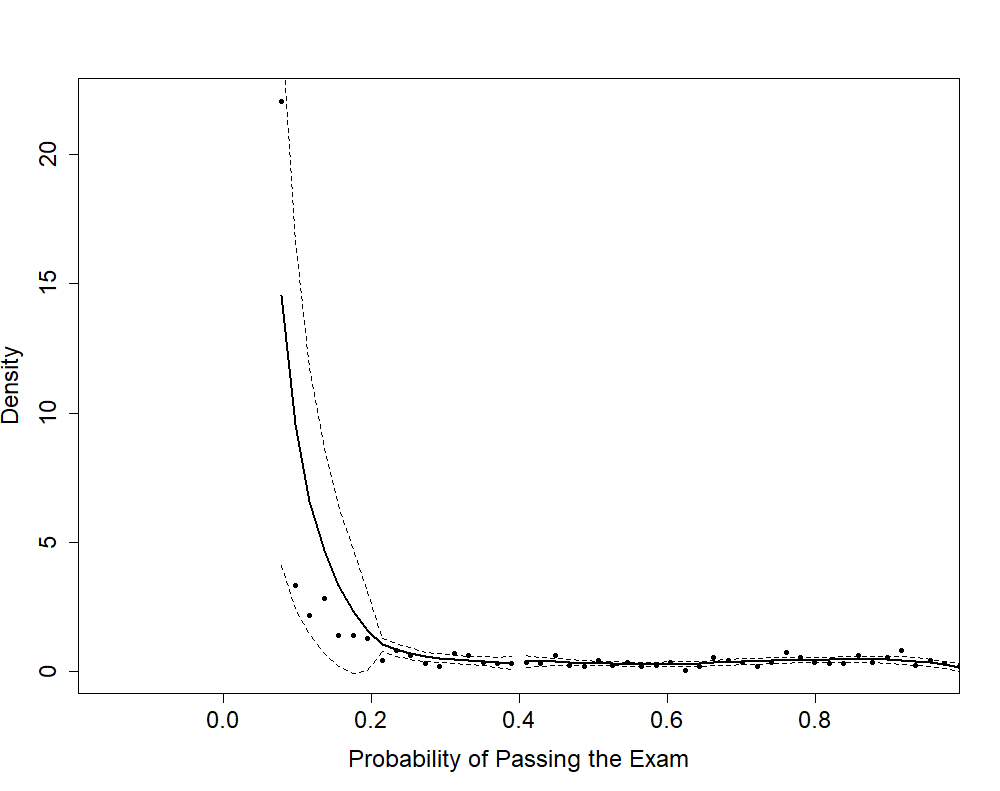}
  		\caption{The McCrary sorting test for the running variable $W$ predicted probability to pass the exam ($x$-axis). There is no jump in the density around the cutoff point of $0.4$, i.e., the density left, and right of the cutoff do not differ substantially.}
	\label{fig:mccrary}
\end{figure}

\section{\uppercase{Empirical Results and Discussion}} \label{sec:analysis}

Table \ref{tab:group_overview} shows that the treatment (warning = $1$) and control (warning = $0$) groups, as expected, differ substantially. The performances in the online tests and JACK score\footnote{Note that students may partially or entirely study outside of the JACK framework. However, since the final exam was taken via JACK, students have a strong incentive to mainly learn on the platform to get used to the framework.}, are much lower in the treatment group. This also illustrates that a single OLS regression of $Y_i$ on $T_i$ would fail to identify the effect of the intervention. 

We first check that the assumption (i) of a continuous running variable with no jump at the cutoff is met. For this, we perform the \cite{mccrarymanipulation2008} sorting test, which tests the continuity of the density of our running variable $W$ -- the predicted probability of passing the exam -- around the cutoff. In order to estimate the effect $\alpha$ correctly, there must not be a jump in the density at the cutoff. Otherwise, some participants could have manipulated the treatment, and the results would no longer be reliable.

The McCrary sorting test indicates no discontinuity of the density around the cutoff with a p-value of $0.509$; see Figure~\ref{fig:mccrary}. Since there is no jump around the cutoff and the students were not informed beforehand about the warning email, we can be relatively confident that the students were not able to manipulate the treatment. In any case, the incentive to worsen one's own performance to receive the treatment seems rather small as there is no direct benefit from receiving the warning. The idea and a possible effect of the email lie in a change in the effort students invest from that time on.

\begin{figure}[]
	\centering
  		\includegraphics[width=0.5\textwidth]{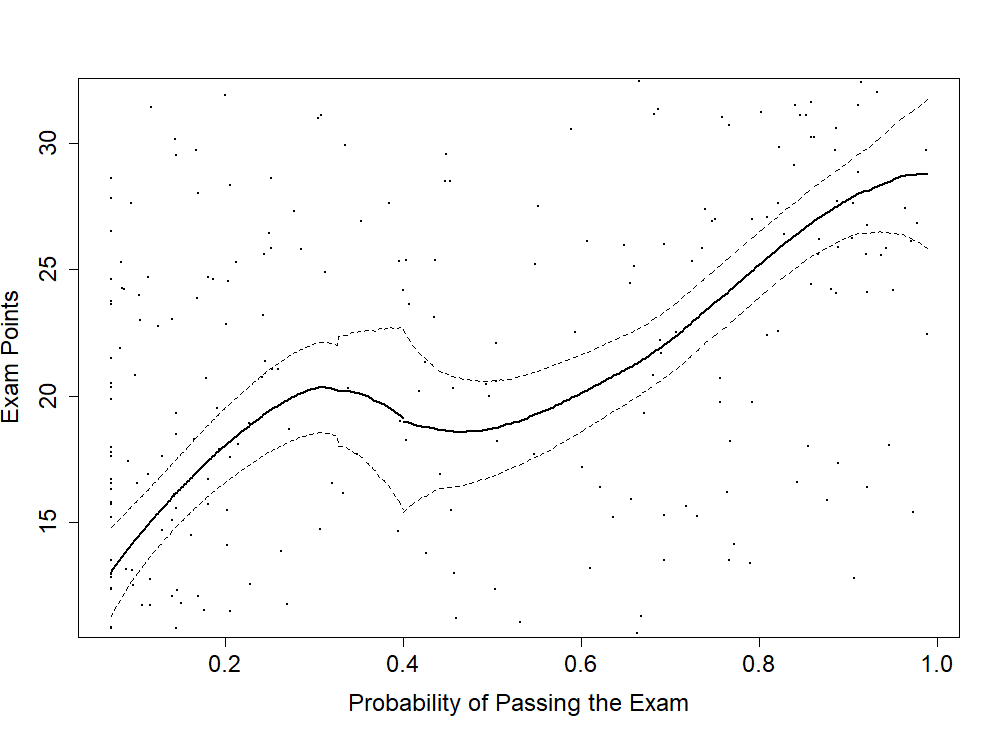}
  		\caption{Graphical illustration of the RDD with the probability to pass the exam $W$ on the $x$-axis and the exam points $Y$ on the $y$-axis. At the cutoff point of $c = 0.4$, we cannot see any (major) decrease in earned exam points.}
	\label{fig:mod}
\end{figure}

\begin{table}[t]
\centering
		\caption{Model 1 summarises the RDD model \textit{without} and Model 2 \textit{with} covariates. LATE describes the Local Average Treatment Effect. The bandwidth and $F$-statistics report the \cite{imbensoptimal2009} bandwidth and the F-statistics for the non-parametric estimation. $N$ is the number of observations used. 
} 
\begin{tabular}{@{}lcc@{}}
\toprule
                                                              & Model 1                                                 & Model 2                                                 \\ \midrule
 LATE                                                   & \begin{tabular}[c]{@{}c@{}}0.193\\ (4.889)\\ \end{tabular} & \begin{tabular}[c]{@{}c@{}}0.146\\ (4.852)\end{tabular} \\
 \midrule
Bandwith                                                      & 0.255                                                   & 0.255                                                   \\
$F$-statistics   $\qquad$                                                & 0.257                                                   & 0.257                                                   \\ 
$N$                                                             & 126                                                     & 126                                                     \\
\bottomrule

\multicolumn{3}{l}{\begin{tabular}[c]{@{}l@{}} \small \textit{Note:} Standard errors are indicated within parenthesis.\\ 
$^{\ast} p < 0.1$, $^{\ast \ast} p < 0.05$, $^{\ast \ast \ast} p < 0.01$.
\end{tabular}}
    
\end{tabular}
\label{tab:mods}
\end{table}

We employed the non-parametric local average treatment effect (LATE) method to estimate the causal effect. This RDD method only compares the local average around the cutoff ($c = 0.4$) rather than fitting a polynomial regression.\footnote{We also performed the RDD using polynomial regression. The results were essentially identical to those of the LATE method.} This method is more efficient than polynomial regression since the estimation involves fewer parameters. Furthermore, we avoid the common concerns of fitting polynomial regressions, e.g., determining the polynomial degree and the tendencies to extremes at the edges \cite{Lee2010}. We performed all RDD regressions with and without covariates, which is usually not necessary in the sharp setting but is often recommended in the fuzzy design to get a more precise estimate of the treatment effect. Including covariates can increase the explained variance of the model \cite[p. 515]{Huntington-Klein2022}.  

The effect point estimates (LATE) in Table~\ref{tab:mods} of the two non-parametric RDD models are positive but not significant (parameter $\alpha$ in equation \eqref{eq:2_stage}). Figure~\ref{fig:mod} gives a graphical illustration of the model.  The LATE is $0.193$ for the estimation without covariates (model $1$ in Table~\ref{tab:mods}) and $0.146$ if we include the covariates (model $2$ in Table~\ref{tab:mods}), with corresponding standard errors of $4.889$ and $4.852$. An estimate of $0.193$ or $0.146$ means that students who received the warning email achieved $0.193$ or $0.146$ points more than comparable students who did not. Compared to the $60$-point final exam, the effect size seems limited.

The bandwidth of $0.255$ was determined with the data-driven approach of \cite{imbensoptimal2009}. The method fits the bandwidth as widely as possible without introducing other confounding effects, e.g., general ability. Therefore, only students with a predicted probability between $0.145$ and $0.655$ ,$0.4$ (cutoff) $\pm \ 0.255$ (bandwidth), are included in the analysis. This leads to a sample size for the estimation of $126$ students ($N$). Other bandwidths were considered in the estimation process but were too conservative or violated the assumption that the groups around the cutoff must be comparable. We provide further discussion in section \ref{sec:conclusion}.

Hence, our RDD results do not provide evidence that the warning email has a significant effect on the students' results (or behavior). This might have several reasons. For instance, many participants who received a warning did not participate in any final exam. Of the 608 students who received a warning, only 183 sat the exam. This likely compromises the detection of an effect. A possible explanation is that the warning might lead weak students to postpone participation to a later semester. The email could give students the impression that the chances of getting a good grade are already relatively low. Therefore, students might be more likely to repeat the course a year later. In a sense, this can also be viewed as a positive outcome, as we then at least prevent students from collecting malus points, cf.~footnote 2.

The non-parametric method used here has the disadvantage that (many) students, which are relatively far from the cutoff, are not included in the analysis. This reduces the effective sample size, and thus precise estimation of the treatment becomes more difficult. Figure~\ref{fig:mod} suggests another potential issue: both groups' variance around the cutoff ($c = 0.4$) is rather high. 

Another important aspect of this analysis is that students can get feedback on their proficiency JACK homework and earn extra points through the online tests and the Kahoot!~games. From the perspective of the students, this is probably an even bigger incentive than the warning email. Hence, the incremental effect of the warning email may be limited. Further incentives and feedback combined with the warning emails during the semester, might have a greater effect on student performance than the warning emails alone. 

\section{\uppercase{Conclusions}}\label{sec:conclusion}

In this paper, we analyzed whether students who perform relatively poorly in a current course can be positively influenced by a warning email. Even though we did not find a significant effect of the treatment, we see the open and transparent communication of the student's  performance to the students as a positive aspect of the system. Furthermore, we are considering expanding the system further. One possible approach we are currently viewing is implementing an automatic repeatedly system for detecting inactive students or students whose submissions show little progress, which will regularly notify students at risk. 

One more aspect that requires attention in the future is the investigation of a possible effect of the warning system on the dropout rate, i.e., whether the email leads to more students withdrawing from the exam beforehand and thus increasing the pass rate. However, a higher dropout rate is not inherently negative. Students can focus on other courses and thus achieve higher grades.

To conclude, we do not detect a significant effect of the warning emails in our design. This is still noteworthy because the successful motivation of weak and modest students remains challenging for instructors.

\section*{Acknowledgments}
We thank all colleagues who contributed to the course ``Induktive
Statistik'' in the summer term 2019.

Part of the work on this project was funded by the German Federal
Ministry of Education and Research under grant numbers 01PL16075 and 01JA1910 and by the Foundation for Innovation in University Teaching under grant number FBM2020-EA-1190-00081.

%
\bibliographystyle{abbrv}
\bibliography{own.bib}  
%
%

\end{document}